# Leveraging Test Driven Development with Large Language Models for Reliable and Verifiable Spreadsheet Code Generation: A Research Framework

Simon Thorne, Cardiff School of Technologies, Cardiff Metropolitan University

Advait Sarkar, Microsoft Research Cambridge, University of Cambridge & UCL

Sthorne@cardiffmet.ac.uk, Advait@Microsoft.com

**ABSTRACT**

*Large Language Models (LLMs), such as ChatGPT, are increasingly leveraged for generating both traditional software code and spreadsheet logic. Despite their impressive generative capabilities, these models frequently exhibit critical issues such as hallucinations, subtle logical inconsistencies, and syntactic errors, risks particularly acute in high stakes domains like financial modelling and scientific computations, where accuracy and reliability are paramount. This position paper proposes a structured research framework that integrates the proven software engineering practice of Test-Driven Development (TDD) with Large Language Model (LLM) driven generation to enhance the correctness of, reliability of, and user confidence in generated outputs. We hypothesise that a "test first" methodology provides both technical constraints and cognitive scaffolding, guiding LLM outputs towards more accurate, verifiable, and comprehensible solutions. Our framework, applicable across diverse programming contexts, from spreadsheet formula generation to scripting languages such as Python and strongly typed languages like Rust, includes an explicitly outlined experimental design with clearly defined participant groups, evaluation metrics, and illustrative TDD based prompting examples. By emphasising test driven thinking, we aim to improve computational thinking, prompt engineering skills, and user engagement, particularly benefiting spreadsheet users who often lack formal programming training yet face serious consequences from logical errors. We invite collaboration to refine and empirically evaluate this approach, ultimately aiming to establish responsible and reliable LLM integration in both educational and professional development practices.*

## 1. INTRODUCTION

The rise of Large Language Models (LLMs) has introduced powerful new paradigms for software and end user development. With their ability to generate syntactically and often semantically correct code, LLMs such as ChatGPT, Claude, and DeepSeek have permeated professional development environments, spreadsheet systems, and “no code” platforms. Yet despite their fluency, these models remain fundamentally unreliable. Hallucinations, subtle logic errors, off by one mistakes, and the generation of plausible but incorrect formulas or programming outputs are common even when tasks appear relatively straightforward (Liu et al 2024, Wang et al 2024, Zhang et al, 2025a, Zhang et al 2025b). This unreliability makes LLMs risky in contexts requiring accuracy, reproducibility, or compliance, such as financial modelling, scientific computation, or safety critical systems.

This problem is particularly acute in spreadsheets, widely used by non-programmers to express complex logic with significant real-world consequences. Prior research has established the serious risks associated with spreadsheet errors and the difficulty of verifying formula correctness, especially in large or opaque models. The recent integration of LLMs into spreadsheet platforms (e.g., Microsoft Excel’s Copilot) presents both opportunities and challenges: users can now specify complex logic in natural language, automatically receiving generated formulas, yet they may lack the expertise or tools necessary to critically evaluate those outputs.

In response, this paper proposes a structured research framework investigating whether Test Driven Development (TDD), a methodology from professional software engineering, can effectively complement LLMs to enhance the reliability, correctness, and usability of generated code and formulas. TDD involves a disciplined, test first approach, wherein developers specify the expected behaviour of a program by creating tests prior to implementing any code. We hypothesise that integrating this structured approach with LLM driven code generation will serve as both a technical constraint and a cognitive scaffold, encouraging the production of more precise and verifiable outputs across diverse programming contexts, including spreadsheet logic, scripting languages such as Python, and strongly typed languages such as Rust.

An additional dimension of our framework explores how this test driven interaction can foster the development of key computational thinking skills and prompt engineering competencies, encouraging users to explicitly articulate the intended behaviour and scope of solutions, a critical skill for effective AI interaction. We adapt the scope of Wing's concept of computational thinking (Wing, 2006); in our context we take it to mean the set of mental processes enabling individuals to break down complex problems systematically, design effective abstractions, create precise tests to verify solutions, and iterate solutions effectively. Complementarily, we define prompt engineering as the structured practice of crafting precise, clear, and strategically targeted prompts to elicit the desired outputs from generative AI systems. Prompt engineering involves iterative refinement, structured reasoning, and clarity in specification skills increasingly recognised as critical for effectively leveraging AI assistance (Denny et al., 2024).

By embedding TDD principles into interactions with LLMs, participants may develop clearer, test oriented prompts, potentially enhancing their overall computational thinking by encouraging them to explicitly articulate the intended behaviour and scope of solutions. Thus, our study seeks to assess whether combining TDD and LLMs not only improves immediate outputs but also enhances users' long term capabilities in computational reasoning and effective AI interaction.

This is a position paper rather than an empirical report. Our objective is to articulate the motivation, scope, and structure of a forthcoming study designed to empirically evaluate the effectiveness and broader educational implications of integrating TDD with LLM assisted code and formula generation.

Specifically, we describe:

- The rationale for applying TDD principles to guide LLM driven development.
- The selection of tasks and domains for testing, explicitly including spreadsheet scenarios.
- The design of experimental groups, comparing LLM only and combined TDD+LLM conditions.
- Evaluation metrics addressing correctness, testability, user comprehension, hallucination rates, computational thinking, and prompt engineering effectiveness.
- Illustrative example tasks demonstrating how structured, test first reasoning can shape LLM interactions toward more robust and verifiable outputs.

### 1.1 Hypotheses

We hypothesise that the combination of Test Driven Development and GPT based assistance will lead to measurable improvements in key aspects of programming practice and cognitive development. Specifically, we expect this combination to foster enhanced computational thinking, particularly in code decomposition and test case design, while also improving code comprehension and error diagnosis. Furthermore, we anticipate that the structured discipline of TDD will support more effective and strategic use of GPT, cultivating better prompting behaviours such as clear specifications and test-oriented language. Finally, we expect participants using TDD+GPT to experience greater confidence, a stronger sense of control, and deeper engagement in the development process, indicative of more meaningful learning and cognitive investment.

- **H1:** The use of TDD in combination with GPT will enhance participants' computational thinking skills, particularly in code decomposition and test case design, compared to GPT alone.
- **H2:** Participants using TDD + GPT will demonstrate improved code comprehension and error diagnosis skills, as measured by debugging and code modification tasks.
- **H3:** TDD + GPT participants will report higher confidence, perceived control, and engagement during the development task, reflecting deeper learning and cognitive engagement.
- H4: Participants in the TDD+GPT group will demonstrate more strategic prompting behaviours, including clearer specifications and test oriented language, indicating development of prompt engineering as a computational skill.

As prompting skill is increasingly recognised as critical to effective AI use (Denny et al., 2024), our experiment indirectly captures the development of prompting strategy in both groups.

### 1.2 Research contributions

This research contributes to both the software engineering and computing education literature by investigating how Test Driven Development (TDD) can serve as a pedagogical scaffold when used in conjunction with Large Language Models (LLMs). While previous studies have demonstrated that LLMs can generate correct code in many scenarios, concerns remain regarding their impact on learning, code comprehension, and cognitive engagement, especially in educational settings where understanding, not just output, is the goal.

Our first contribution is to explore whether combining TDD with LLM based code generation improves not only the correctness of outputs but also the user's comprehension of the logic and structure of those outputs. This addresses a key concern raised in recent GenAI pedagogy literature (Kazemitabaar et al., 2025; Denny et al., 2024), which emphasises that students may passively accept AI generated code without engaging in the deeper reasoning required for genuine understanding.

Second, we frame TDD as a form of metacognitive scaffolding, a structured process that externalises the programmer's reasoning through test construction, helping learners plan, monitor, and evaluate their problem solving strategies. Prior work (Edwards, 2003; Denny et al., 2019) supports the idea that test first approaches encourage reflection and decomposition, core elements of computational thinking. Our study design will examine whether this scaffolding effect applies when TDD is used not with human authored code, but with LLM assisted development.

Third, we contribute a practical experimental framework for testing how different approaches to code generation, LLM only versus LLM + TDD, affect learner outcomes, including comprehension, confidence, and the ability to debug or adapt existing code. The framework includes a full pretest, experimental coding task, post test comprehension measure, and structured GPT prompting logs, making it extensible for future studies.

Finally, this work offers a novel lens on prompt engineering as a teachable skill within the broader domain of computational literacy. By requiring participants to construct test cases before engaging with GPT, we encourage the development of precise, test oriented prompts linking LLM usage to structured reasoning practices essential in both professional development and computer science education.

## 2.0 Related work

The impact of Generative AI on teaching practices, the effectiveness of various learning interventions, and innovations in tools encompasses a vast literature, which we shall not aim to reproduce here. Prather et al. (2025) provide a comprehensive overview. Germane to our concern of test driven development as a potential GenAI forward methodology of instruction are their findings from interviews of tool creators and educators studying and using GenAI. These include the changing nature of assessment, shift in emphasis towards learning process rather than correct answers, changing in programming competencies (code reading, testing, decomposition, debugging become more important than writing code from scratch), and the need to train students to use GenAI in industry.

Below, we shall aim to focus on the most relevant narrative, which is on the intersection of how these tools affect the nature and practice of programming, and how instruction can be delivered most effectively to students.

As Lau and Guo (2023) find, programming instructors are divided as to the appropriate pedagogical response to the rise of code generating LLMs. Instructors were focused on preventing AI assisted cheating and divided on whether to ban AI use entirely or to integrate them meaningfully to prepare students for the future world of programming work.

## 2.2 Improving learning outcomes from code authoring

Much prior research has been focused on investigating how code authoring tasks can be modified to take advantage of LLMs while retaining learning outcomes. Since LLMs can produce correct code, students risk losing the benefits of the manual process of writing correct code and its attendant cognitive and learning outcomes. Thus, much research focuses on alternative modes of introducing LLMs into the coding process, abstaining from the direct provision of correct code as much as possible. An early example of this is Code Help by Liffiton et al. (2023), which provides on demand AI generated help for student queries, providing explanations and guidance, identifying incomplete or ambiguous queries and prompting the student for additional information, but never including complete solutions that can be copied and pasted. While learning outcomes were not directly evaluated, students qualitatively found value in the responses provided by Code Help.

Hou et al. (2024) evaluated CodeTailor, which provides concise and personalised Parsons puzzles to help students with coding tasks; the rationale is to provide support by generating puzzle pieces that could produce a correct program, but still maintaining learning and cognitive engagement by requiring students to understand enough about the code and the solution to solve the puzzle. While post test scores were no greater than when students practiced using a baseline (which simply used an LLM to generate a correct answer that could be copied and pasted without cognitive engagement), the authors find evidence that practice with Code Tailor could improve student understanding of specific coding concepts (e.g., dictionary keys). Nonetheless, they found that Parsons puzzles did not carry sufficient explanatory power to help students learn specific details, and that Code Tailor could also generate solutions that exceeded learners' current knowledge.

Kazemitabaar et al. (2024) evaluated CodeAid, an educational programming assistant designed to provide helpful responses (such as answering conceptual questions, generating pseudo code with line by line explanations, and annotating incorrect student code with suggestions for fixes) without revealing code solutions. Their analysis finds that AI coding assistants for students should support students in making the decision to use the AI tool and in formulating queries, be balanced in terms of directness and learning engagement, and support trust, transparency an control.

Following on from this study, Kazemitabaar et al. (2025) evaluated a number of alternatives for improving student engagement and comprehension of AI generated code, finding that "lead and reveal" (AI guides students through step by step problem solving before revealing code) and "trace and predict" (students trace the code line by line to follow the values of variables) as being the most balanced at inducing learning without incurring overwhelming cognitive load.

### 2.3 Novel pedagogical approaches to programming / CS education with GenAI

While most approaches thus far have focused on modifying or mediating code authoring tasks with a view to improving student engagement and thus comprehension and learning outcomes, relatively few have directly targeted the issue that in practice, as working programmers, students will have access to correct code generating language models, and that coding exercises and teaching methods could be adapted with this in mind. We review these below.

Denny et al. (2023) propose that the effective prompting of language models is a skill that ought to be developed and propose "prompt problems" which are direct exercises in prompting language models to produce code. In a variation, or perhaps expansion of this method, Denny et al. (2024) also propose exercises that require students to explain code in plain English (EiPE) until the output of a code generating LLM based on their explanation becomes equivalent to the original code being explained, on the basis that such exercises target both the development of the student's code comprehension and their prompting skills simultaneously.

Jin et al. (2024) take the idea further and propose using language models as teachable agents, adopting a learning by teaching framework, in which students learn programming by having a conversation in which they teach programming concepts to a chatbot (roleplaying as a programming novice), finding that this approach improved the "knowledge density" of conversations without a significant effect on cognitive load, but did not have an effect on improving metacognition.

### 2.4 Test driven development for CS education

Research in CS education has recognised the value of testing   developing suitable tests, as well as solving example test cases (i.e., working through a problem statement with a concrete example) before code is written   for its potential to provide "metacognitive scaffolding" (Denny et al., 2019). Indeed, as early as 2003, Edwards (2003) proposed that CS education be rethought from a test first perspective. This resulted in a pedagogical approach known as test first programming (e.g., see Erdogmus et al., 2005).

However, the research on applying TDD as an instructional method mostly predates the introduction of commercially available coding agents; work remains to be done to establish its efficacy and its participation in this new regime, particularly as correct code generation from natural language prompts (at least for the complexity of programming typically required in introductory courses) is now a "solved" problem.

### 2.5 Test driven development as GenAI forward development methodology

The application of generative AI to the test driven development methodology in general (i.e., not specifically in education) has begun to be explored; we are not the first to propose a fit between TDD and LLM based code generation (e.g., see Mathews and Nagappan, 2024, who demonstrate that including test cases leads to higher success rates for LLM based code generation on benchmarks).

Much recent work has focused on the potential of GenAI to produce appropriate test cases. This may be fully automated, e.g., TestPilot (Schäfer et al., 2023) and ChatUniTest (Chen et al., 2024), or involve a human in an interactive and guided process of intent formalisation, e.g., TiCoder (Fakhoury et al., 2024). As another example, Mock et al. (2024) provide preliminary evidence that GenAI can be useful both when used solely for code generation (with human written tests) and also for writing tests themselves, though the authors stress that expert supervision is still required to ensure the quality of the produced code.

Piya and Sullivan (2024) provide some best practices for test driven development using large language models based on empirical experience applying ChatGPT to LeetCode problems using TDD, such as sanitising descriptive function names, providing unit tests avoiding input output pairs, representing unique string manipulations earlier in the test set, providing descriptive naming for tests, etc.

## 3.0 METHODOLOGY

### 3.1 Introduction

The following section explores the methodological strategy of the investigation, detailing the experimental design, the characteristics of the participants, the means of recruitment, the materials used, the data collection infastructure and a detailed description of the resarch procedure.

### 3.2 Experimental Design

The research will used a controlled experiment that will make comparisons between one group using TDD and GPT in combination and the other group only using GPT. Participants are randomly assigned to one of two clearly controlled experimental groups:

Group 1 (GPT Only):

- Participants use GPT without structured testing guidelines.
- Participants are explicitly advised they may optionally write tests, but this is not mandatory.

Group 2 (TDD + GPT):

- Participants explicitly follow TDD methodology.
- Participants must explicitly write tests first, implementing code to pass tests using GPT assistance.

In both groups, participants will document their GPT interactions by copying their entire GPT sessions (prompts and responses) into a provided Word document, enabling detailed qualitative analysis of interactions.

### 3.3 Participants

We aim to recruit approximately 40-50 participants drawn from the second year “Concepts in AI” module at Cardiff Metropolitan University. The participants will come from BSc Computer Science and Software Engineering programmes hence these participants should have a good basic understanding of foundational computing concepts and generative AI, ensuring baseline coding proficiency suitable for the tasks and familiarity with generative AI. In addition, the experiment will be conducted towards the end of the term, allowing the students to benefit from the contents of the module which examines, symbolic AI reasoning, machine learning approaches and generative AI. Of course there will be some variation, Individual levels of experience will be captured in the pre-test .

#### 3.3.1 Recruitment Method

Participants will be recruited directly through class announcements, email invitations, and institutional messaging platforms, clearly stating the voluntary nature of participation.

**Inclusion Criteria:**

- Basic knowledge of programming fundamentals (validated via pre-test).
- Basic Knowledge of LLMs, Generative AI and ChatGPT (validated via pre-test)

**Exclusion Criteria:**

- Individuals lacking basic programming proficiency.
- Individuals unable to commit fully to the two-hour experimental duration.

### 3.4 Materials & Setup

Participants will use the following tools and resources clearly defined for this study:

#### 3.4.1 Software and Coding Environment:

- ChatGPT 4 via OpenAI's web interface for both groups
- Python programming language within Jupyter labs or participants' preferred IDEs.
- GitHub Copilot or similar AI coding assistants explicitly excluded to maintain experimental clarity.

While our study focuses on code-based development rather than spreadsheets directly, we selected Python as the primary programming language due to its conceptual and structural similarities with spreadsheets. Python offers a clear, readable syntax that aligns well with formula-based reasoning and cell-based computation. Importantly, Python is increasingly adopted by spreadsheet users seeking greater analytical flexibility, automation, and scalability, particularly through tools like pandas that mirror spreadsheet operations in code (McKinney, 2012; Microsoft, 2021).

#### 3.4.2 Data Collection Infrastructure:

- **Qualtrics:** Administered pre-test and post-test surveys to capture quantitative and qualitative data on participant skills, confidence, and experience.
- **OneDrive:** Secure HTTPS server storage for all participant data, including anonymised code submissions, ChatGPT session logs, and survey responses.

### 3.5 Procedure

Participants will follow a clearly defined experimental workflow:

#### 3.5.1 Step 1: Pre-test Assessment (15 20 minutes)

The study will be conducted using Qualtrics and will assess participants' programming skills through a basic coding task, such as implementing a vowel-counting function, and a set of multiple-choice questions covering Python fundamentals, including topics like time complexity and built-in methods. Participants' familiarity and confidence with Test-Driven Development (TDD) will be evaluated by asking them to write explicit test cases for a given problem. In parallel, their familiarity with GPT will be assessed through a practical GPT-assisted coding task, such as creating a function to calculate an average. Given the variability in participants' baseline programming ability, TDD experience, and GPT familiarity, we will stratify assignment to conditions based on these demographic and skill-related factors, rather than use pure random allocation. This stratification approach is particularly important in the context of a small sample size, to ensure balanced group comparisons and reduce confounding effects.

Example pre-test task

> *"Write a Python function to count the vowels in a string. Example input: 'hello world'. Expected output: 3."*

#### 3.5.2 Step 2: Experimental Coding Task (90 minutes)

Participants are randomly allocated into two clearly defined experimental groups:
**GPT Only Group:**

- Participants will use GPT freely to implement a specified coding task (task management system).
- Testing is optional and at the participant's discretion.

**TDD + GPT Group:**

- Participants explicitly follow TDD principles, using the "Red (failing test) → Green (implement feature) → Refactor" cycle.
- GPT explicitly assists in generating code snippets, debugging failed tests, and suggesting additional test cases.

Prior to the task, TDD + GPT participants will receive a brief instructional sheet outlining the core rules and stages of TDD, which will be reinforced through in-class and offline teaching materials. The guidance will emphasise writing tests first, implementing only the necessary functionality to pass those tests, and then refactoring for clarity or efficiency.

#### 3.5.3 Main coding task:

The participants of either group will be asked to create a simple but comprehensive task management system, a to do list with all the usual features that you would expect of such a system.

> *"Develop a simple task management system capable of adding, editing, marking as complete, deleting tasks, and flagging overdue tasks based on their due dates."*

This system should allow users to add new tasks with associated details such as descriptions and due dates, edit existing task details, mark tasks as complete, and delete tasks as needed. Additionally, the system will include functionality to automatically identify and flag tasks that are overdue based on the current date and their assigned due dates. The goal is to assess participants' ability to manage data structures, handle user input, implement basic date and time logic, and design an intuitive interface or clear workflow. This exercise will also evaluate participants' understanding of practical application development, their coding efficiency, and their capacity to integrate various programming concepts to deliver a cohesive solution.

#### 3.5.4 Step 3: Post test Survey and Modification Task (10 15 minutes)

The post-test will be conducted via Qualtrics and will capture several key metrics. Participants' code comprehension will be evaluated through a self-assessment rating on a Likert scale (1 5) and through their responses to a written debugging scenario, where they must identify and resolve a hypothetical coding bug. Developer experience will be measured using Likert scale feedback, explicitly capturing participants' perceived ease of task completion, the helpfulness of GPT and TDD methodologies, and qualitative feedback highlighting any challenges encountered during the tasks. Additionally, participants will complete a code modification task, specifically implementing a small enhancement involving adding priority levels to tasks and sorting them, accordingly, thereby directly assessing adaptability and deeper comprehension of the provided codebase.

### 3.6 Analysis plan

A mixed-methods approach will be used to evaluate the effects of combining Test-Driven Development (TDD) with GPT on programming performance, comprehension, prompting behaviour, and learner confidence.

**Quantitative Analysis**

- Pre/post comparisons within each group will use paired t-tests (or non-parametric equivalents) to assess improvements in programming accuracy, test design, and comprehension.
- Between-group comparisons will use independent t-tests or Mann-Whitney U tests to evaluate:
    - Code quality and feature completeness
    - Debugging and code modification accuracy
    - Likert-scale measures of confidence, control, and perceived usefulness
- Prompt quality scoring will assess the use of test-oriented language, clarity, and structure. Scores will be compared across groups to evaluate strategic prompting (H4).

**Qualitative Analysis**

- Thematic coding of GPT session logs will identify patterns in prompting, code generation, debugging, and test-related behaviours.
- Open-ended survey responses will be analysed to capture perceived challenges, strategies, and reflections on GPT and TDD.

**Triangulation**

- Findings from code artefacts, prompt logs, and survey responses will be triangulated to provide a holistic view of how TDD and GPT influence learning, strategy, and coding behaviour.

### 3.7 Conclusions on Methodology

This methodology builds on research demonstrating that TDD supports systematic thinking and reduces omission errors, particularly in novice programming contexts (Beck, 2003; George & Williams, 2004). In the context of computer science education, where learners often default to reactive or incomplete coding strategies, the structure imposed by TDD offers a valuable scaffold for reasoning about program behaviour before implementation. While the experiment itself has not yet been conducted, the following section provides illustrative examples to demonstrate how TDD test cases can be operationalised into both Python functions and equivalent Excel formulas. These examples serve to clarify the experimental intent and highlight the kinds of reasoning and representational translation that participants will be expected to perform.

## 4.0 PRACTICAL EXAMPLES OF TDD FOR PROGRAMMING AND SPREADSHEETS

This section demonstrates how TDD can be combined with GPT to generate both Python code and Excel formulas. The test cases are presented, given to GPT which then generates Python and Excel statements. It includes four common spreadsheet tasks: compound interest calculations, date-based logic, inventory restocking, and tiered pricing logic. For each, we present test cases, a TDD style Python implementation, and an equivalent Excel formula.

These examples are intended only for showing the process by which code can be generated from TDD and as an indicator of the quality of code such a process can yield. All examples have been passed through ChatGPT4o, the prompt used simply contains the TDD cases and the instruction to generate both Python and Excel code.

### 4.1 Interest Calculation

Compound interest calculations often suffer from subtle logic errors, table 1 contains TDD cases.

| Test Case | Principal (£) | Rate (%) | Years | Expected Result (£) |
|---|---|---|---|---|
| TC1 | 1,000 | 5 | 2 | 1,102.50 |
| TC2 | 500 | 3 | 5 | 579.64 |
| TC3 | 2,000 | 4 | 3 | 2,249.73 |
| TC4 | 0 | 5 | 10 | Invalid input |

**Table 1 TDD test cases for interest calculation**

**Python**

```
def compound_interest(principal, rate, years):
    if principal < 0 or rate < 0 or years < 0:
        return "Invalid input"
    return round(principal * (1 + rate / 100) ** years, 2)
```

**Excel**

```
=IF(OR(A1<0, B1<0, C1<0), "Invalid input", ROUND(A1 * (1 + B1 / 100) ^ C1, 2))
```

Where: A1 = Principal, B1 = Interest Rate, C1 = Years

### 4.2 Overdue Task Flagging in Project Management

Date comparisons in spreadsheets are prone to logic errors, table 2 shows test cases for flagging overdue tasks in a project management application.

| Test Case | Due Date | Today's Date | Expected Result |
|---|---|---|---|
| TC1 | 01/01/2025 | 02/01/2025 | Overdue |
| TC2 | 01/01/2025 | 01/01/2025 | On Time |
| TC3 | 02/01/2025 | 01/01/2025 | On Time |
| TC4 | TODAY()   1 | TODAY() | Overdue |

**Table 2 TDD test cases for task flagging**

**Python**

```
from datetime import date, timedelta

def task_status(due_date, today=None):
    if today is None:
        today = date.today()
    return "Overdue" if due_date < today else "On Time"
```

**Excel Formula**

*=IF(TODAY()>A1, "Overdue", "On Time")*

Where: A1 contains the due date.

Or alternatively:

*=IF(A1 < B1, "Overdue", "On Time")*

### 4.3 Inventory Restocking Logic

Managing stock levels involves logic to determine whether restocking is needed, table 3 contains some suitable TDD cases.

| Test Case | Current Stock | Reorder Threshold | Expected Result |
|---|---|---|---|
| TC1 | 50 | 100 | Restock |
| TC2 | 120 | 100 | Sufficient Stock |
| TC3 | 100 | 100 | Sufficient Stock |
| TC4 | 0 | 10 | Restock |

**Table 3 TDD test cases for stock reorder**

**Python Code (TDD)**

```
def restock_needed(current_stock, reorder_threshold):
   return "Restock" if current_stock < reorder_threshold else "Sufficient Stock"
```

**Excel Formula**

*=IF(A1 < B1, "Restock", "Sufficient Stock")*

Where: A1 = Current Stock, B1 = Reorder Threshold

### 4.4 Tiered Discount Pricing

This example introduces conditional pricing logic where discounts depend on purchase amounts, see table 4.

| Test Case | Order Total (£) | Expected Final Price (£) |
|---|---|---|
| TC1 | 50.00 | 50.00 |
| TC2 | 100.00 | 95.00 |
| TC3 | 499.99 | 474.99 |
| TC4 | 500.00 | 450.00 |
| TC5 | 999.99 | 899.99 |
| TC6 | 1000.00 | 850.00 |
| TC7 | 0.00 | 0.00 |
| TC8 | 100.00 | Invalid input |

**Table 4 TTD cases for discounted pricing logic**

**Python**

```
def discounted_price(order_total):
    if order_total < 0:
        return "Invalid input"
    elif order_total < 100:
        return round(order_total, 2)
    elif order_total < 500:
        return round(order_total * 0.95, 2)
    elif order_total < 1000:
        return round(order_total * 0.90, 2)
    else:
        return round(order_total * 0.85, 2)
```

**Excel**

```
=IF(A1<0, "Invalid input",  IF(A1<100, A1,  IF(A1<500, ROUND(A1*0.95, 2),
IF(A1<1000, ROUND(A1*0.90, 2),  ROUND(A1*0.85, 2)))))
```

Where: A1 contains the total order value.

## 5.0 DISCUSSION

While TDD has its origins in professional software engineering, the structured, example first mindset it promotes may be especially suitable for end user developers particularly those working with spreadsheets. Unlike traditional programming environments, spreadsheet logic is often constructed in an ad hoc fashion, making it difficult to validate correctness or catch subtle errors. There is also a lack of testing in spreadsheet applications, meaning that errors are made but not discovered until the mistake is realised. The result is a particularly high error rate in spreadsheets (Panko, 2008) with errors of omission representing the most common type of mistake made. By articulating test cases that cover all required system behaviors, including edge cases and negative scenarios, developers are more likely to notice missing logic or unimplemented functionality thus potentially reducing the rate at which omission errors are made.

We also feel that this combination could be particularly impactful for spreadsheet developers wishing to leverage LLMs in their end user programming activities. There have been many attempts to port software engineering principles and techniques to spreadsheets for instance, the use of Data Flow Diagrams (DFDs) and other diagrammatic methods for planning logic and system structure (Rajalingham, 2001; Hermans, 2011; Mireault, 2015), development methodologies tailored for spreadsheet modelling (Grossman & Ozluk, 2010), and even efforts to transform Excel into a structured programming environment, such as Model Master (Paine, 2001, 2005). These approaches were motivated by the desire to increase rigor, reduce errors, and improve maintainability.

However, such efforts have historically had mixed uptake. One must remember that most spreadsheet developers are not trained computing professionals, and nor are they interested in becoming them. They are domain experts working in finance, operations, education, or science, who use spreadsheets as a flexible modelling tool. Hence, any technique offered as an improvement must be *palatable* and *immediately consumable* by such users. Tools that impose additional abstraction or formality often risk rejection.

By contrast, the TDD + GPT method may offer a more lightweight, conversational, and approachable route to improved rigor. It does not require users to learn new formal languages or diagrammatic systems. The “test first” approach introduces a gentle layer of discipline that helps make reasoning explicit without demanding software engineering expertise (Wilson et al 2003, Rothermel et al. 1998)

Recent research in CS education has emphasised structured, scaffolded use of LLMs to support learning without displacing key cognitive processes. Denny et al. (2019) highlight test driven programming as a metacognitive aid, while Kazemitabaar et al. (2025) find that stepwise guidance (“lead and reveal”) balances engagement and cognitive load. Tools like CodeAid and CodeTailor also show that conversational, structured AI use can support student agencies without handing over solutions. Our TDD + GPT approach aligns with these trends, offering a lightweight, prompt-based structure that supports reasoning and reduces hallucinations particularly for spreadsheet users who benefit from accessible, low- barrier methods.

In this sense, our proposed framework suggests a *cognitive bridge* between natural language goals and structured logic. It combines the power of LLMs with the scaffolding of test-driven thinking, potentially offering a method that is both robust and accessible particularly for spreadsheet developers seeking to enhance reliability without leaving their familiar tools and practices behind.

## 5.0 CONCLUSION AND FURTHER WORK

This position paper has proposed a research framework for evaluating whether combining Test Driven Development (TDD) with LLM based code generation can serve as both a technical constraint and a cognitive scaffold. Our aim is not only to reduce hallucinations and logical errors but to explore how first thinking can improve code comprehension, confidence, and prompting discipline.

By treating TDD as a metacognitive structure and combining it with LLMs’ generative fluency, we offer a hybrid methodology that speaks to both educational and professional goals. The proposed experiment, drawing on structured task design and pre/post analysis, seeks to illuminate how learners interact with GPT-based tools in both code and spreadsheet environments.

We invite collaboration from educators, developers, and HCI researchers to test, extend, and adapt this framework. If effective, it could offer a scalable model for responsible LLM use in education, coding bootcamps, and even industry settings, where correctness and clarity are paramount, but LLM output remains opaque.

Going forward, our immediate next steps involve implementing the experimental framework outlined here. This experiment will test our core hypotheses and provide insights into the effectiveness of combining TDD with LLM-based code and spreadsheet generation. The study will establish clear benchmarks and success indicators through structured tasks, pre/post analysis, and reflective feedback.

Future research should also investigate longitudinal impacts, examining whether sustained TDD-LLM integration fosters enduring improvements in programming habits and critical thinking. Additionally, expanding the scope of inquiry to include comparative analyses across different LLM architectures and prompting strategies will deepen our understanding of optimal practices.

Ultimately, this work seeks to contribute broadly to the ongoing dialogue on integrating AI responsibly within technical education and professional development. We look forward to engaging with a diverse community to refine and evolve this framework collaboratively.

## REFERENCES

Anaconda, Inc. (2020). *The state of data science 2020: Moving from hype toward maturity*. https://www.anaconda.com/state-of-data-science-2020

Chen, Y., Hu, Z., Zhi, C., Han, J., Deng, S., & Yin, J. (2024, July). Chatunitest: A framework for llm based test generation. In Companion Proceedings of the 32nd ACM International Conference on the Foundations of Software Engineering (pp. 572 576).

Denny, P., Leinonen, J., Prather, J., Luxton Reilly, A., Amarouche, T., Becker, B. A., & Reeves, B. N. (2024, March). Prompt Problems: A new programming exercise for the generative AI era. In Proceedings of the 55th ACM Technical Symposium on Computer Science Education V. 1 (pp. 296 302).

Denny, P., Prather, J., Becker, B. A., Albrecht, Z., Loksa, D., & Pettit, R. (2019, November). A closer look at metacognitive scaffolding: Solving test cases before programming. In Proceedings of the 19th Koli Calling international conference on computing education research (pp. 1 10).

Denny, P., Smith IV, D. H., Fowler, M., Prather, J., Becker, B. A., & Leinonen, J. (2024). Explaining code with a purpose: An integrated approach for developing code comprehension and prompting skills. In Proceedings of the 2024 on Innovation and Technology in Computer Science Education V. 1 (pp. 283 289).

Edwards, S. H. (2003, October). Rethinking computer science education from a test first perspective. In Companion of the 18th annual ACM SIGPLAN conference on Object oriented programming, systems, languages, and applications (pp. 148 155).

Erdogmus, H., Morisio, M., & Torchiano, M. (2005). On the effectiveness of the test first approach to programming. IEEE Transactions on software Engineering, 31(3), 226 237.

Fakhoury, S., Naik, A., Sakkas, G., Chakraborty, S., Musuvathi, M., & Lahiri, S. (2024, April). Exploring the effectiveness of llm based test driven interactive code generation: User study and empirical evaluation. In Proceedings of the 2024 IEEE/ACM 46th International Conference on Software Engineering: Companion Proceedings (pp. 390 391).

Grossman, T., & Ozluk, O. (2010). Spreadsheets Grow Up: Three Spreadsheet Engineering Methodologies for Large Financial Planning Models. Proceedings of the 11th annual conference of the European Spreadsheets Risks Interest Group (EuSpRIG). London. doi:https://arxiv.org/abs/1008.4174

Hermans, F. (2011). Breviz: Visualizing Spreadsheets using Dataflow Diagrams. Proceedings of the 11th annual conference of the European Spreadsheets Risks Interest Group. doi:https://doi.org/10.48550/arXiv.1111.6895

Hou, X., Wu, Z., Wang, X., & Ericson, B. J. (2024, July). Codetailor: Llm powered personalized parsons puzzles for engaging support while learning programming. In Proceedings of the Eleventh ACM Conference on Learning@ Scale (pp. 51 62).

Jin, H., Lee, S., Shin, H., & Kim, J. (2024, May). Teach ai how to code: Using large language models as teachable agents for programming education. In Proceedings of the 2024 CHI Conference on Human Factors in Computing Systems (pp. 1 28).

Kazemitabaar, M., Huang, O., Suh, S., Henley, A. Z., & Grossman, T. (2025, March). Exploring the design space of cognitive engagement techniques with ai generated code for enhanced learning. In Proceedings of the 30th International Conference on Intelligent User Interfaces (pp. 695 714).

Kazemitabaar, M., Ye, R., Wang, X., Henley, A. Z., Denny, P., Craig, M., & Grossman, T. (2024, May). Codeaid: Evaluating a classroom deployment of an llm based programming assistant that balances student and educator needs. In Proceedings of the 2024 chi conference on human factors in computing systems (pp. 1 20).

Lau, S., & Guo, P. (2023, August). From "Ban it till we understand it" to" Resistance is futile": How university programming instructors plan to adapt as more students use AI code generation and explanation tools such as ChatGPT and GitHub Copilot. In Proceedings of the 2023 ACM Conference on International Computing Education Research Volume 1 (pp. 106 121).

Liffiton, M., Sheese, B. E., Savelka, J., & Denny, P. (2023, November). Codehelp: Using large language models with guardrails for scalable support in programming classes. In Proceedings of the 23rd Koli Calling International Conference on Computing Education Research (pp. 1 11).

Liu, F., Liu, Y., Huang, H., Wang, R., Yang, Z., & Zhang, L. (2024). Exploring and Evaluating Hallucinations in LLM Powered Code Generation. Arxiv. doi:arXiv:2404.00971

Mathews, N. S., & Nagappan, M. (2024, October). Test Driven Development and LLM based Code Generation. In Proceedings of the 39th IEEE/ACM International Conference on Automated Software Engineering (pp. 1583 1594).

McKinney, W. (2012). *Python for data analysis: Data wrangling with Pandas, NumPy, and IPython*. O'Reilly Media.

Microsoft. (2021). *From Excel to Python*. Microsoft Learn. https://learn.microsoft.com/en-us/training/paths/from-excel-to-python/

Mireault, P. (2015). Developing a Repeating Model Using the Structured Spreadsheet Modelling and Implementation Methodology. Proceedings of the 15th annual conference of the European Spreadsheets Risks Interest Group. London. doi:https://arxiv.org/abs/1602.06453

Mock, M., Melegati, J., & Russo, B. (2024, June). Generative AI for Test Driven Development: Preliminary Results. In International Conference on Agile Software Development (pp. 24 32). Cham: Springer Nature Switzerland.

Paine, J. (2001). Ensuring Spreadsheet Integrity with Model Master. Proceedings of the 2nd annual conference for the European Spreadsheets Risks Interest Group (EuSpRIG). Amsterdam. doi:https://arxiv.org/abs/0801.3690

Paine, J. (2005). Exelsior: Bringing the Benefit of Modularisation to Excel. Proceedings of the 5th annual conference of the European Spreadsheets Risks Interest Group (EuSpRIG). London. doi:https://arxiv.org/abs/0803.2027

Panko, R. (2008). Spreadsheet errors: What we know and what we think we can do. Proceedings of 8th annual conference of The European Spreadsheets Risks Interest Group (EuSpRIG), London, doi:https://doi.org/10.48550/arXiv.0802.3457

Piya, S., & Sullivan, A. (2024, April). LLM4TDD: Best practices for test driven development using large language models. In Proceedings of the 1st International Workshop on Large Language Models for Code (pp. 14 21).

Prather, J., Leinonen, J., Kiesler, N., Gorson Benario, J., Lau, S., MacNeil, S., ... & Zingaro, D. (2025). Beyond the Hype: A Comprehensive Review of Current Trends in Generative AI Research, Teaching Practices, and Tools. 2024 Working Group Reports on Innovation and Technology in Computer Science Education, 300 338.

Rajalingham, K. (2001). A revised classification of spreadsheet errors. Proceedings of the 2nd annual conference of the European Spreadsheets Risks Interest Group (EuSpRIG). Amsterdam. Retrieved from https://eusprig.org/wp-content/uploads/rajalingham2005.pdf

Rothermel, G., Li, L., DuPuis, C., & Burnett, M. (1998, April). What you see is what you test: A methodology for testing form-based visual programs. In *Proceedings of the 20th international conference on Software engineering* (pp. 198-207). IEEE.

Rust. A, Bishop. B, McDaid. K, (2006), Investigating the Potential of Test Driven Development for Spreadsheet Engineering, Proceedings of the 6th annual conference of the European Spreadsheets Risks Interest Group (EuSpRIG), Cambridge, https://arxiv.org/abs/0801.4802

Schäfer, M., Nadi, S., Eghbali, A., & Tip, F. (2023). An empirical evaluation of using large language models for automated unit test generation. IEEE Transactions on Software Engineering, 50(1), 85 105.

Wang, L., Deng, X., Wen, H., You, M., Liu, W., Li, Q., & Li, J. (2024). Prompt engineering in consistency and reliability with the evidence-based guideline for LLMs. NPJ Digital Medicine, 7(41). doi:https://doi.org/10.1038/s41746-024-01029-4

Wing, J. M. (2006). Computational thinking. *Communications of the ACM*, *49*(3), 33-35.

Wilson, A., Burnett, M., Beckwith, L., Granatir, O., Casburn, L., Cook, C., ... & Rothermel, G. (2003, April). Harnessing curiosity to increase correctness in end-user programming. In *Proceedings of the SIGCHI conference on Human factors in computing systems* (pp. 305-312).

Zhang, Y., Li, S., Liu, J., Yu, P., Han, C., Mckeown, K., . . . Ji, H. (2025a). The Law of Knowledge Overshadowing: Towards Understanding, Predicting, and Preventing LLM Hallucination. Arxiv. doi:https://doi.org/10.48550/arXiv.2502.16143

Zhang, Z., Wang, Y., Wang, C., Chen, J., Zheng, Z., & Yat-sen, S. (2025b). LLM Hallucinations in Practical Code Generation: Phenomena, Mechanism and Mitigation. Arxiv.